\documentstyle[aps,prl,epsfig]{revtex}

\begin{document}

\title{La-site shift as probe for Jahn-Teller effect}

\author{Bas B. van Aken, Auke Meetsma, and Thomas T. M. Palstra$^*$\nocite{pal}}
\address{Solid State Chemistry Laboratory, Materials Science Centre, University of Groningen,
\\ Nijenborgh 4, 9747 AG  Groningen, the Netherlands}

\date{\today}
\twocolumn[\hsize\textwidth\columnwidth\hsize\csname@twocolumnfalse\endcsname

\maketitle

\begin{abstract}
We observe two consecutive transitions in La$_{1-x}$Ca$_x$MnO$_3$, $x=0.19$: ferromagnetic ordering at $T_c=177$ K and Jahn-Teller orbital ordering at $T_{JT}=150$ K. The presence of a ferromagnetic insulating state below $T_{JT}$ shows that the metallic phase is bound by ferromagnetic and Jahn-Teller ordering, and not induced by a critical doping concentration. The A-site shift is found to be a good indicator of Jahn-Teller ordering. Furthermore, the A-site shift does not disappear above $T_{JT}$, indicating a dynamical behaviour of the distorted MnO$_6$ octahedra.
\end{abstract}

]

The basic interactions in the manganite perovskites allow three phases: a 
ferromagnetic metal, a charge/orbital ordered antiferromagnetic insulator and a 
paramagnetic polaronic liquid. Metallicity is obtained by introducing holes by 
doping in antiferromagnetic, insulating LaMnO$_3$. This doping renders 
La$_{1-x}$Ca$_x$MnO$_3, $for $0.20<x<0.50$, both metallic and ferromagnetic, as 
the interactions are dominated by double exchange. However, 
La$_{1-x}$Ca$_x$MnO$_3$, with $0.10<x<0.20$, has a ferromagnetic insulating 
ground state. This unexpected coexistence of ferromagnetic and insulating 
behaviour seems to contradict the conventional double  and super exchange 
models. Originally the magnetic state was thought to be a canted 
antiferromagnetic phase,\cite{Kaw96b} but experiments determined the magnetic 
state to be ferromagnetic\cite{Ram97}. The origin of the coexistence of 
ferromagnetism with insulating behaviour is not clear, but might stem from a 
delicate balance of charge localisation by orbital ordering (OO), due to the 
Jahn-Teller (JT) effect, and ferromagnetic interactions between Mn$^{3+}$- 
Mn$^{4+}$. The exact position of the phase line of the JT ordering transition 
in the composition temperature phase diagram of doped LaMnO$_3$ is not known, 
and may depend on other variables, such as the tolerance factor and the 
magnetic ordering.

The phase diagram of Sr doped manganites has been explored in great detail. 
Here the situation is more complicated than for Ca doping, because the number 
of phases is larger due to the rhombohedral structure at $x>0.18$ and the 
pronounced charge ordering (CO) at $x\sim1/8$.\cite{Yam96} Several authors 
reported a JT related structural phase transition above the magnetic ordering 
temperature, $T>T_c$ at $x\sim0.12$. Below $T_c$, a transition to CO or OO is 
observed, where the cooperative JT distortion is significantly 
reduced.\cite{Arg96,End99} As the transition temperatures are extremely 
concentration dependent, a  comparison between the various reports is not 
straightforward. It is claimed that the intermediate phase is both 
ferromagnetic and metallic and exhibits static cooperative JT 
distortions.\cite{Kaw96b,End99} Some reports clearly distinguish these two 
properties and combine short range order of JT distortions with metallic 
behaviour.\cite{Dab99} However, a general relation between the JT ordered phase 
and the nature of the conductivity has not been established. Also, a 
coincidence of the CO transition and the re-entrant insulator-metal transition 
is claimed. The common metal-insulator transition is indisputably associated 
with the ferromagnetic ordering at $T_c$. \cite{Kaw96b,End99,Dab99}

The Ca doped phase diagram is somewhat less complex, as there is no 
orthorhombic-rhombohedral structural transition. Furthermore, the phase 
transitions take place at higher concentrations. As a result we can probe the 
ferromagnetic insulating phase at concentrations far away from $x=1/8$ to 
prevent charge ordering. In this Letter we explore the region where the JT 
ordering phase line has crossed the magnetic ordering phase line. We will show 
that the transition to the ferromagnetic metallic phase is not at fixed carrier 
concentration, but is controlled by the suppression of JT ordering. 
Conventionally, the JT ordering is observed via the Mn-O distances. We will 
provide evidence that the La-site shift is a more accurate tool for JT 
ordering, as the La position is sensitive to a change in the oxygen 
environment. Our measurements suggest that above the JT ordering phase line 
both the metallic phase and the paramagnetic phase exhibit strong JT like 
fluctuations. These fluctuations become long range ordered below $T_{JT}$.

The experiments were carried out on single crystals of La$_{1-x}$Ca$_x$MnO$_3$, 
$x=0.19$, obtained by the floating zone method at the MISIS institute, Moscow. 
Although all crystals were twinned,\cite{Van01c} small mosaicity and sharp 
diffraction spots were observed. Furthermore, the sharp magnetic and electronic 
transitions indicate the good quality of the crystals. Simultaneous 
measurements of resistance $R$ and magnetisation $M$ were performed in a MPMS 
magnetometer to find the exact transition temperatures for this composition. A 
thin piece was cut from the crystal to be used for single crystal 
diffractometry. Initial measurements were carried out on an Enraf-Nonius CAD4 
single crystal 4-circle diffractometer to determine the twin fraction 
volume.\cite{Van01c} Temperature dependent measurements between 90 K and 300 K 
were performed on a Bruker APEX diffractometer with an adjustable temperature 
set-up.

\begin{figure}[htb]
\centering
\includegraphics[height=51.2mm,width=70mm]{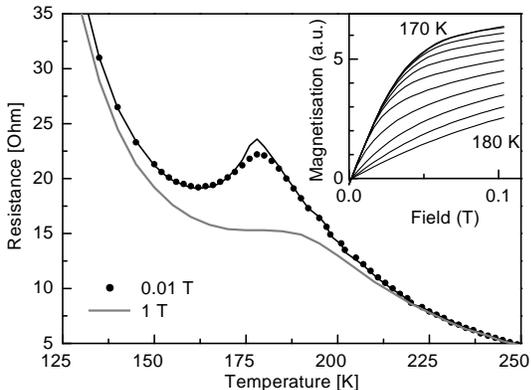}
\caption{Temperature dependence of the resistance at $H=0$, 0.01 and 1 T. The 
inset shows the magnetisation vs. applied magnetic field in temperature steps 
of 1 K between 170-180 K. Both measurements establish that $T_c=177$~K.} 
\label{R(T)}
\end{figure}

In Fig.~\ref{R(T)} the temperature dependence of the resistivity is shown in 
zero, small and large magnetic fields. The resistance has a clear maximum at 
177 K. At significantly lower temperatures, $T\approx160$ K, the resistivity 
shows a subtle and wide transition to activated behaviour. This maximum can be 
suppressed by applying small fields ($H<0.1$ T). Upon applying larger fields 
the resistance decreases not only in the local maximum but in the whole 
temperature range, both below the local minimum at $T\approx160$ K and 
above $T_c$.

In the inset of Fig.~\ref{R(T)}, magnetisation curves are plotted in the 
temperature range $170<T<180$ K with temperature steps of 1 K. For $T\leq177$ K 
the initial slope is constant and determined by the demagnetisation factor. 
Therefore, we establish $T_c$ to be 177 K. Both the magnetisation curves and 
resistance measurements indicate a sharp transition at $177\pm0.5$ K.

\begin{figure}[htb]
\centering
\includegraphics[height=43.7mm,width=75mm]{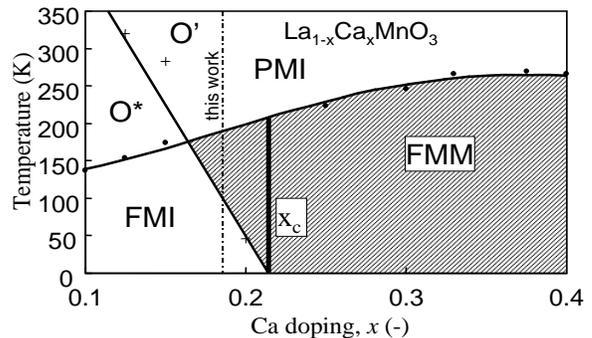}
\caption{Sketched phase diagram in the FMI-FMM transition region, modified from 
Cheong \emph{et al.}.\protect\cite{Che00} The critical concentration, $x_c$, only 
indicates the doping induced insulator to metal transition at $T=0$.} 
\label{phasediagram}
\end{figure}

We propose that the observed upturn in resistance at $T\leq160$ K is a result 
of re-entrant insulating behaviour caused by JT ordering. The well-known phase 
diagram by Cheong \emph{et al.}\cite{Che00} is modified as shown in 
Fig.~\ref{phasediagram}. We propose that the FMM phase, hatched area, should be 
extended to the phase line, which indicates the JT ordered to JT disordered 
transition. Obviously, JT ordering and metallicity are mutually exclusive. In 
analogy to conventional ferromagnetic metallic La$_{1-x}$Ca$_x$MnO$_3$ systems, 
with $x\sim0.3$,\cite{Boo98} we expect to see a narrowing of the distribution 
of Mn-O bond lengths below $T_c$ as a result of the itinerancy in the 
ferromagnetic, metallic regime. As soon as the JT orbital ordering sets in 
there will be a separation of the Mn-O bond lengths, as observed for 
La$_{1-x}$Sr$_x$MnO$_3$ with $0.11<x<0.165$.\cite{Dab99} The difference in bond 
lengths should become less pronounced in the charge ordered phase at the lowest 
temperatures.\cite{Arg96,Kaw96b,Dab99}

From our full structure refinement, we can clearly observe the difference in 
in-plane Mn-O bond lengths, {\it i.e.} Mn-O2, both in the paramagnetic phase 
and in the insulating phase. However, we can not observe a decrease in this 
difference in the metallic regime, $160<T<177$ K. The error bars on the bond 
lengths are quite large, due to the relatively low X-ray scattering factor of 
oxygen and the influence of the twinning, common in many perovskite 
materials.\cite{Veg86,Rod98} As we explained in detail elsewhere, the 
reflections of the various twin fractions overlap indistinguishably, which 
averages the calculated $\sqrt{2}a$, $b$ and $\sqrt{2}c$ 
parameters.\cite{Van01c} Therefore the accuracy of the lattice parameter 
determination is not as good as for neutron powder diffraction.

However, the refinement of the relative atomic positions within the unit cell, 
which are reflected in the observed intensities, is extremely accurate. 
Therefore it makes more sense to focus on the refined atomic positions instead 
of the bond lengths. This holds especially for the O2 (in-plane) oxygen 
position, since it completely determines the Jahn-Teller distortion. For the 
full structure determination, we had to derive the twin relations, which are 
reported elsewhere.\cite{Van01c} The temperature dependence of the O2 
fractional coordinates show a transition in $x_{O2}$, but this provides little 
insight into the physical mechanism. 

\begin{figure}[htb]
\centering
  \includegraphics[height=51.1mm,width=60mm]{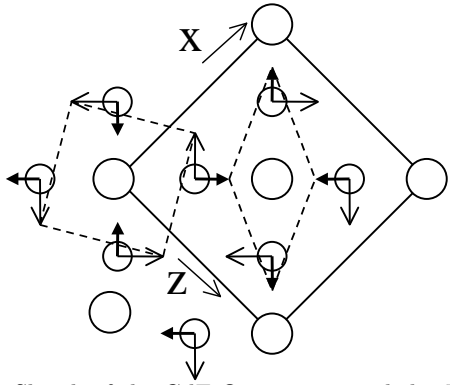}
\caption{Sketch of the GdFeO$_3$ rotation and the JT distortion in the 
$ac$-plane, obeying $Pnma$ symmetry. Mn and O are represented by large and 
small circles, respectively. The shift associated with JT is shown as a closed 
arrow (shift along $x+z$). Open arrows indicate the GdFeO$_3$ rotation, with a 
shift along $x-z$. }\label{plane}
\end{figure}

More insight is gained by using the parameters $x+z$ and $x-z$ as sketched in 
Fig.~\ref{plane}. Here, a movement of the O2 ion parallel to $x-z$, keeping 
$x+z=\frac{1}{2}$, results in equal bond lengths but a Mn-O-Mn angle smaller 
than 180{$^\circ$}. We interpret this movement, along $x-z$, as the GdFeO$_3$ 
rotation. Similarly, a shift of the O-ion along $x+z$, fixing $x-z=0$, results 
in different in-plane bond lengths and therefore indicates a Jahn-Teller 
distortion. The undistorted cubic structure obeys $x+z=\frac{1}{2}$ and $x-z=0$.

\begin{figure}[htb]
\centering
  \includegraphics[height=52.3mm,width=75mm]{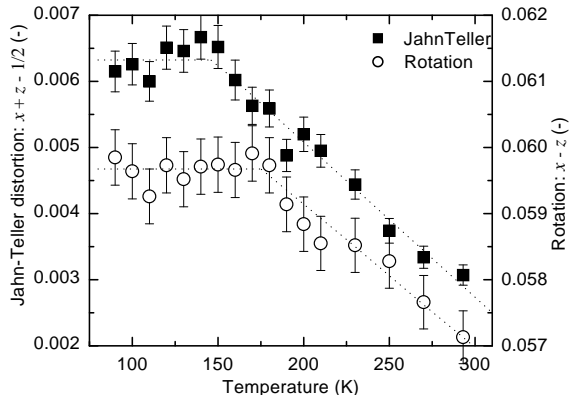}
\caption{Jahn-Teller distortion and GdFeO$_3$ rotation as a function of 
temperature. The JT-distortion increases linearly down to $T_{JT}=150$ K. A 
kink in the GdFeO$_3$ rotation appears at $T=180$ K.}\label{JT}
\end{figure}

The measures for the GdFeO$_3$ rotation and the Jahn-Teller distortion are 
plotted in Fig.~\ref{JT}. We see a clear increase in the Jahn-Teller parameter 
from the room temperature phase, $\approx0.003$, to the low temperature phase 
with $\approx0.006$. It increases continuously down to $T_{JT}=150$ K. The 
rotation increases slightly with decreasing $T$ and levels off at $T\sim180$ K. 
This small increase is associated with a freezing of the octahedra at low 
temperature.

\begin{figure}[tb]
\centering
  \includegraphics[height=47.9mm,width=75mm]{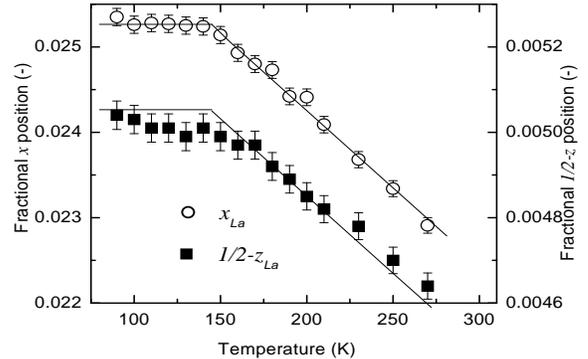}
\caption{$x_{La}$ and $z_{La}$ positions vs. $T$. Drawn lines are linear, 
$T>150$ K, and constant, $T<150$ K, fits of $x_{La}$. For the $z_{La}$ a fixed 
ratio of 1:5 with respect to $x_{La}$ is assumed.} \label{Laxz}
\end{figure}

Recently, Mizokawa {\it et al.} reported on the interplay between the GdFeO$_3$ 
rotation, the orbital ordering and the A-site shift in ABO$_3$, with B=Mn$^{3+} 
(3d^4)$ or V$^{3+} (3d^2)$. Their theoretical calculations suggest that the 
observed orbital ordering in LaMnO$_3$ is stabilised by both a large GdFeO$_3$ 
rotation and a shift of the A-site ion.\cite{Miz99} Conversely, if a 
Jahn-Teller distortion is present, then the energy will be lowered if it is 
accompanied by a shift of the A-site. In Fig.~\ref{Laxz}, we show the 
temperature dependence of the A position, ($x_{A}, \frac{1}{4}, z_{A}$), with 
respect to the ideal position ($0 \frac{1}{4} \frac{1}{2}$). Note that the 
error bars and scatter are much smaller than for the O positions, due to the 
higher electron density at the A-site.

Fig.~\ref{Laxz} shows a linear increase of $x_A$ with $T$ down to $T_{JT}=150$ 
K. $x_A$ is roughly 5 times larger than $z_A$ for all temperatures. Below 
$T_{JT}$, $x_A$ and $z_A$ are temperature independent. This La-site shift along 
[1 0 5] is in good agreement with the [1 0 7] direction which was assumed by 
Mizokawa {\it et al.}.\cite{Miz99} Furthermore, the observed temperature 
dependence is in excellent agreement with the temperature dependence of the JT 
effect.

\begin{figure}[htb]
  \centering  
  \includegraphics[height=47.9mm,width=75mm]{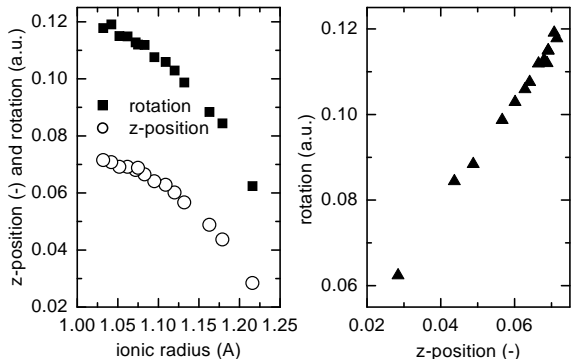}
\nopagebreak\caption{Left panel: The rotation and $z_A$ versus the ionic radius 
in AFeO$_3$. Both parameters indicate a deviation from the cubic perovskite, 
and both increase with decreasing $r_A$. Right panel: The rotation versus 
$z_A$. The parameters show almost perfect correlation indicating their intimate 
relation.}
  \label{rot-zet}
\end{figure}

The basic argument favouring the A-site shift is that the covalency between A 
and O can be optimised by decreasing the A-O distances for the three shortest 
bonds. Any GdFeO$_3$ distortion will result in a distorted polyhedron and 
therefore it will cause an A-site shift. Marezio {\it et al.} have studied the 
structure of the AFeO$_3$ compounds in great detail and their data allows us to 
focus on the correlation between the rotation and the La-site 
shift.\cite{Mar70} The relevant atomic parameters, $x$ and $z$ of both the O2 
and the A-site, are all fully correlated as shown in Fig.~\ref{rot-zet}. The 
figure shows a perfect linear relation between the A-site shift and the 
rotation parameter. Thus the shift of the A-site atom is, in this system, fully 
determined by the rotation of the FeO$_6$ octahedron. A fixed rotation will 
result in a constant A-site shift. An extra shift of the A-site at fixed 
rotation must therefore indicate the presence of a further influence on the 
oxygen positions. In the La$_{1-x}$Ca$_x$MnO$_3$ system the extra influence is 
the ordering of the Mn$^{3+}$ $e_g$ orbitals. The observed increase of 0.0027 
in the rotation corresponds to an increase of 0.0021 in the La-site shift, 
using the relation between rotation and A-site shift. The observed shift is 
much larger. 

We interpret our data as follows. Between 90 and $T_{JT}=150$ K the average Mn 
environment is distorted, and the $e_g$ orbitals are ordered in the d-type 
fashion.\cite{Miz99} Due to the orbital ordering, the charge carriers are 
localised and the material behaves as an insulator. We have not observed any 
super lattice reflections. Any charge ordering phase either exists at lower 
temperatures, $T<90$ K, or at a hole concentration closer to $x=1/8$. In the 
phase diagram of La$_{1-x}$Sr$_x$MnO$_3$, the CO phase borders the FMM phase, 
as observed by superlattice reflections in single crystal neutron 
experiments.\cite{Yam96} In contrast, for La$_{1-x}$Ca$_x$MnO$_3$ the CO phase 
is suppressed by the orbital ordered FMI phase.

Above $T_JT=150$ K the La-site shift and the parameter for JT ordering begin to 
decrease. The rotation remains constant, as does the magnetic ordering. One 
expects equal Mn-O distances in the metallic, itinerant phase.\cite{Boo98} The 
absence of equal bond lengths indicates that the structure on average is not a 
fully itinerant phase, although the JT parameter decreases. The low 
conductivity in the metallic regime agrees well with this interpretation. 
However, the decrease in the JT parameter signals the destruction of long range 
orbital ordering, although medium-range correlations remain. This orbital 
ordering melting is sufficient to render the material metallic, though the 
metallicity is not associated with the absence of Jahn-Teller distortions, only 
with the absence of long range orbital order. The Bragg peaks signal this order 
because the time scale of diffraction is very small, $\sim10^{-15}$ s. A 
decrease of the integrated intensity can thus be attributed to a decrease in 
phase coherence on length scales  $100-1000$ \AA. The decrease of the La-site 
shift and the JT parameter indicates that the long range orbital ordering is 
broken. Locally the distortions are still present, but ordered on smaller 
length scales, in good agreement with experiments probing the local 
distortions.\cite{Bil96} This implies that metallic behaviour and long range JT 
ordering can not coexist in the same phase, in contradiction to the 
interpretation of \cite{Kaw96b,End99}. 

Above $T_c$, we observe no further change in the structure, except the 
continuous decrease of the La-site shift and the JT parameter. The magnetic 
ordering is broken, which decreases the 'bare' electron kinetic energy and the 
conduction becomes semiconducting as the electron-phonon coupling is the 
leading term.\cite{Mil98} The effect of the magnetic ordering on the structure 
is smaller than can be observed via single crystal X-ray diffraction. The 
magnetically ordered state allows metallic conduction, but the mobility is 
impeded by JT fluctuations. Eventually, the breaking of the magnetic ordering 
leads to a semiconducting state, with localised charge carriers due to the JT 
fluctuations. With increasing temperature the average structure will have less 
and less the signature of the JT ordered phase. 

We have demonstrated that the ferromagnetic metallic phase is obtained, in a 
limited temperature range, by the suppression of the long range Jahn-Teller 
ordering. This contrasts with the common opinion that metallicity occurs if the 
charge carrier density exceeds a critical concentration. Furthermore, we have 
shown that we can study the Jahn-Teller ordering by observing the La-site 
shift. The JT ordering is no longer long range above $T_{JT}$, but on shorter 
length scales the JT distortions do not disappear. With increasing temperature 
there is a simultaneous reduction of the JT parameter and the La site shift. 
The metallic state of La$_{1-x}$Ca$_x$MnO$_3$ is bound by ferromagnetic 
ordering and the absence of orbital ordering.

Stimulating discussions with Daniel Khomskii, Lou-F\'{e} Feiner, George 
Sawatzky, Graeme Blake, Martine Hennion and Takashi Mizokawa are gratefully 
acknowledged. This work is supported by the Netherlands Foundation for the 
Fundamental Research on Matter (FOM).

\end{document}